\documentstyle[preprint,aps,epsfig]{revtex}

\begin{document}

\tighten

\preprint{} 
 
\title{ Analytical Results for Multifractal Properties of Spectra 
of Quasiperiodic Hamiltonians near the Periodic Chain}

\author{Andreas R\"udinger$^{*}$ and Cl\'ement Sire$^{+}$}

\address{Laboratoire de Physique Quantique (URA 505 du CNRS)\\ 
Universit\'e Paul Sabatier, 118 route de Narbonne, \\ 
31062 Toulouse C\'edex, France}

\maketitle

\begin{abstract}
The multifractal properties of the electronic spectrum of a general
quasiperiodic chain are studied in  first order in the quasiperiodic potential
strength. Analytical expressions for the generalized dimensions are found and
are in good agreement with numerical simulations. These first order results
do not depend on the irrational incommensurability. 

\end{abstract}

\vskip2cm

 J.\ Phys.\ A {\bf 29} (1996), 3537-3544

\newpage
\section{Introduction}
Quasicrystals are characterized by remarkable transport properties (e.g. low
conductivities $increasing$ with $T$ or disorder) which are generally
attributed to the complicated interplay between the electrons and the peculiar
atomic structure \cite{avignon}. Some of these have been related to
the electronic properties found in 2D or 3D theoretical models (critical
states, spiky singular density of states, anomalous transport...) \cite{avignon}. One dimensional
models of quasiperiodic structures share all these features, in addition to
other properties as the occurrence of a zero measure spectrum. Thus, despite
their lack of direct physical interest, they are worth studying since
properties which can only be illustrated numerically in higher dimensions, and
thus sometimes lack a direct justification, can be calculated very precisely
in 1D.

A frequently studied one-dimensional quasiperiodic tight-binding Hamiltonian 
is given by 
\begin{equation}
H = \sum_i t_{i+1}|i\rangle\langle i+1| + 
v_i |i\rangle\langle i| + t_i |i\rangle\langle i-1|, 
\end{equation}
where the $t_i \in \{1,\rho\}$ and $v_i \in \{-\lambda, \lambda\}$
follow quasiperiodic sequences to be specified below. 
The spectra of this kind of Hamiltonians have been proven to be 
Cantor sets of zero Lebesgue measure for all values of the 
potential $1-\rho \neq 0$, respectively $\lambda \neq 0$ \cite{Suto,Belli1}.
Furthermore, the density of states (DOS) is purely singular continuous
\cite{Suto,Belli1}.  In order to characterize the intricate nature of these
spectra, a multifractal analysis of the measure can be elaborated, yielding a
multifractal spectrum $f(\alpha)$.  More precisely, $f(\alpha)$ is the
Hausdorff dimension of the set of energies in the spectrum with local scaling
exponent $\alpha$, where $\alpha$ characterizes the DOS singularities
($D(E+\Delta E) - D(E) \sim \Delta E^{\alpha}$). The existence of the
generalization of such exponents to 2D and 3D can be shown to directly affect
the transport properties, and a phenomenological transport theory of
quasicrystals involve the distribution of scaling exponents. \cite{avignon}.
In the thermodynamic formalism, introduced by Halsey {\it et al.}, generalized
dimensions $D_q$ can be defined \cite{Halsey,Paladin}.  These also play an
important role for
the understanding of anomalous diffusion properties and the decay of the
autocorrelation function encountered in quantum dynamics of quasiperiodic
systems \cite{Wilkinson,Guarneri,Geisel,Ketzmerick}.
In the case of a quasiperiodic chain with the golden mean as
incommensurability, 
a renormalization analysis based on the trace map approach leads to the exact 
minimal and maximal scaling exponents $\alpha_{\rm min}$ and 
$\alpha_{\rm max}$ for all values of the 
quasiperiodic potential, while the fractal dimensions $f(\alpha)$, 
$\alpha_{\rm min} < \alpha < \alpha_{\rm max}$, remain unknown 
\cite{Ostlund,Kohmoto1,Kohmoto2,Kohmoto3}. 
In the limit of large potential ($|1-\rho| \gg 1,\; |\lambda| \gg 1$) the 
renormalization approach of Niu and Nori\cite{N&N} has been used to obtain 
$f(\alpha)$ \cite{Zheng,Fred} for this Fibonacci chain.

In the present paper we discuss the opposite limit ($|1-\rho| \ll 1,\;
|\lambda| \ll 1$), for which a perturbation theory approach has been
introduced in \cite{Sire} for $any$ quasiperiodic chain. 
We begin by reporting briefly on this approach (II), exploit the results 
in order to calculate the multifractal properties of the energy spectrum (III) 
and finally compare our theoretical relations with numerical data (IV).

\section{Perturbation theory for small potential}

We consider the tight binding model,
\begin{equation}
  t_{i+1}\psi_{i+1}+v_{i}\psi_{i}+t_{i}\psi_{i-1} =E\psi_{i},  
\label{equ:tight-binding}
\end{equation}
with the hopping elements $t_i$ arranged in a quasiperiodic manner: 
\begin{equation}
   t_{i} = \left\{ \begin{array}{cc}
\rho &  \mbox{for} \quad 0<\{\omega i\}<1-\omega, \\
1 & \mbox{for} \quad 1-\omega < \{\omega i\} < 1,
                   \end{array}
        \right.  
\end{equation}
where $\{\cdot \}$ denotes the fractional part and $\omega \in (0,1)$ is an
irrational number.  The onsite potentials are $v_i=\lambda$ if the site $i$ is
surrounded by two  different bonds, and $-\lambda$ otherwise.  In the
following we will consider rational approximants 
with $p_{l}$ hopping elements of strength $\rho$ and $q_{l}$ ones of strength 
1 (i.e. $\omega_l = 1/(1+p_l/q_l)$) and
eventually take the limit $l \to \infty$.  For these approximants with period
$n_l=p_l + q_l$ Bloch theorem can be applied, and one obtains $n_l$ energy
bands.
 
For small $|\lambda|$ and $|1-\rho|$ perturbation theory is used, taking as 
a starting point the nonmodulated  chain ($\lambda = 1-\rho = 0$) with its  
eigenvalues,   
\begin{equation}
E_j^{(0)}(k) = 2\cos\left(\frac{2\pi j + k}{n_l}\right), \; j=0,\cdots,n_l-1,
\end{equation}
where $k$ represents the Bloch vector.
If the quasiperiodic potential is switched on, gaps open at the  
energies $E_j^{(0)}(0) \quad 
(j \neq 0,\; j \neq \frac{n_l}{2} \;(\mbox{for $n_l$ even}))$ 
and $E_j^{(0)}(\pi) \quad (j \neq \frac{n_l+1}{2} \; (\mbox{for $n_l$
  odd}))$. 
For odd values of $n_{l}$ these energies can be renumbered in a compact form: 
\begin{equation}
  E_{j,0} = \pm 2\cos\left(\frac{\pi j}{n_{l}}\right), \quad j =
  1,\cdots,\frac{n_{l}-1}{2}. 
\label{equ:Eper}
\end{equation}
Since in the limit of large system size, the spectral properties will not
depend  on whether $n_l$ is even or odd, we will restrict ourselves, 
for sake of simplicity, to odd values of $n_l$. 

In order to facilitate perturbation theory and to index the gaps roughly 
according to their width, Sire and Mosseri \cite{Sire} 
introduced a new numbering, where the atoms of the quasiperiodic chain are 
numbered with respect  
to their location in perpendicular space rather than with respect to 
their position in physical space. 
In this basis the originally tridiagonal Hamiltonian with quasiperiodic entries
becomes a multidiagonal Hamitonian with identical entries in each secondary
diagonal, except for the two Bloch entries. 

By applying perturbation theory for degenerate states to this Hamitonian it
can be shown that the band edges read in first order \cite{Sire,remark}:   
\begin{equation}
\begin{array}{cc}
E_j \pm \Delta E_j = &2 \epsilon \cos(\pi j \beta_l)(1-(1-\rho)\beta_l) 
+ \lambda(4\beta_l-1) \\
&\pm \frac{2}{\pi j}((1-\rho)\sin(\pi j \beta_l) - \epsilon 
 \lambda \sin(2\pi j \beta_l)), \end{array}
\label{equ:start}   
\end{equation}
with $j \in [1,\cdots,\frac{n_l-1}{2}]$, 
$\epsilon = \pm 1$ and $\beta_l = p_l/n_l = 1-\omega_l$.
In principle, this expression correctly describes the opening of gaps with
index $j<\pi^2/(|\lambda| + |1-\rho|)$. 
For $\lambda= 1- \rho = 0$ all gaps are closed and $E_{j} = 2\epsilon \cos(\pi
j p_{l}/n_{l})$, which is a mere renumbering of equation (\ref{equ:Eper}). 
Equation (\ref{equ:start})  is the starting point for our 
calculation of the multifractal properties.

\section{Multifractal properties of the energy spectra}

In the thermodynamic formalism the multifractal scaling function 
$\tau(q)$ is obtained by requiring that the partition function, 
\begin{equation}
\Gamma(\tau,q) = \frac{1}{n_l^q} \sum_{j=1}^{n_l} \frac{1}{\Delta_j^{\tau}}
\label{equ:Gamma}
\end{equation}
be constant for $n_l \to \infty$, where $\Delta_j$ denotes the widths of the
individual (nonoverlapping) bands. 
For a periodic chain it is straightforward to obtain that  
\begin{equation}
  q^{(0)}(\tau) = \left\{ 
\begin{array}{cc}
\tau +1 & \mbox{for} \; \tau <1, \\
2\tau  & \mbox{for} \; \tau \ge 1, 
\end{array}
\right. 
\end{equation}
so that the multifractal spectrum $f^{(0)}(\alpha)$, 
which is the Legendre transform of $\tau^{(0)}(q)$, consists of two parts: 
$f^{(0)}(1)=1$, corresponding to the absolutely continuous component of the 
DOS, and $f^{(0)}(1/2)=0$, reflecting the presence of van Hove singularities
at the edge of the spectrum. 

In the following, our aim is to calculate the corrections to $q^{(0)}(\tau)$,
respectively $f^{(0)}(\alpha)$, in first order of the quasiperiodic
potential.  We will separately consider two cases: the onsite model
($\rho=1,\; \lambda  \neq 0$) and the off-diagonal model ($\rho \neq 1,\;
\lambda = 0$).  We begin by treating the former model, which is, as we will
see, the more interesting one. 
In order to calculate the multifractal properties by means of perturbation 
theory, we will interpret changes in
the scaling exponent as logarithmic corrections in the size of the system
$n_l$ for $q(\tau)$, in the limit of small perturbation
($n_l^{q}\sim  n_l^{q^{(0)}+\lambda q^{(1)}+\cdots}\sim
n_l^{q^{(0)}}(1+\lambda q^{(1)} \ln n_l + \cdots)$). 
This method is reminiscent of what is achieved in the field of critical
phenomena in order to calculate critical exponents in perturbation theory
(e.g. $\varepsilon$ expansion).

The bandwidths $\Delta_j$ of the modulated chain can be expressed by 
\begin{equation}
  \Delta_j = \Delta_{j,0} - \delta \Delta_j,
\label{equ:Delta}
\end{equation}
where $\Delta_{j,0}$ stands for the corresponding width for the periodic 
chain:
\begin{equation}
\Delta_{j,0} = 2 \left| \cos\left(\frac{\pi j p_l}{n_l}\right) -
\cos\left(\frac{\pi (j p_l+1)}{n_l}\right) \right| = \frac{2\pi}{n_l} \left|
\sin\left(\frac{\pi j p_l}{n_l}\right)\right| + {\cal O}(\frac{1}{n_l^{2}}), 
\end{equation}
and $\delta \Delta_j >0$ denotes the decrease of the bandwidth due to the 
quasiperiodic potential, which can be  expanded  in a formal series, 
\begin{equation}
  \delta \Delta_j = \sum_{l=1}^{\infty} \lambda^l \; \delta \Delta_j^{(l)}. 
\label{equ:exp_q}
\end{equation}
For simplicity we take $\lambda >0$, so that $\delta \Delta_j^{(1)} >0$; 
the final result only depends on $|\lambda|$, which can be easily checked. 
Inserting  equation (\ref{equ:exp_q}) in (\ref{equ:Delta}) and
(\ref{equ:Gamma}) 
we find the partition function in first order in $\lambda$:
\begin{equation}
\Gamma(\tau,q) = \frac{1}{n_l^q}\sum_j \frac{1}{\Delta_j^{\tau}} = 
\Gamma_{0}(\tau,q)\left(1 + \lambda \tau \frac{\displaystyle \sum_j
\Delta_{j,0}^{-1-\tau}\; \delta \Delta_j^{(1)}}{\displaystyle \sum_j
\Delta_{j,0}^{-\tau}} \right),  
\label{equ:exp_Gamma} 
\end{equation}
where $\Gamma_{0}(\tau,q)$ is the partition function of the periodic chain.  
On the other hand, inserting $q(\tau)=\sum \lambda^l q^{(l)}(\tau)$ 
into the asymptotic relation $\Gamma(\tau,q) = n_l^{q(\tau)-q}$ yields 
in first order 
\begin{equation}
\Gamma(\tau,q) = \Gamma_0(\tau,q)(1 + \lambda q^{(1)}(\tau) \ln n_l). 
\end{equation}
Thus the first order correction $q^{(1)}(\tau)$ reads 
\begin{equation}
  q^{(1)}(\tau) = \tau \cdot \lim_{n_l \to \infty} \frac{1}{\ln n_l}\cdot 
\frac{\displaystyle \sum_j \Delta_{j,0}^{-1-\tau} \delta \Delta_j^{(1)}}{
\displaystyle \sum_j \Delta_{j,0}^{-\tau}}.
\label{equ:q1}
\end{equation}
In order to evaluate this expression we remark that the decrease $\delta
\Delta_{j} = \Delta_{j,0} - \Delta_{j}$ of each bandwidth is the sum of the
half-widths of the two adjoining gaps, i.e. in first order:  
\begin{equation}
  \lambda \: \delta \Delta_{j}^{(1)} = \Delta E_{j_{1}} + \Delta E_{j_{2}}. 
\end{equation}
Since $\Delta_{j,0}$ depends smoothly on the location of the band, the sum in
the numerator of equation (\ref{equ:q1}) can be rearranged: 
\begin{equation}
\sum_j \Delta_{j,0}^{-1-\tau}\; \delta \Delta_j^{(1)} = 
\sum_j \Delta_{j,0}^{-1-\tau}\; \frac{2}{\lambda}|\Delta E_j|. 
\end{equation}
In the diagonal case the half-width of the gap is given by 
\mbox{$|\Delta E_j| = \frac{2}{\pi j} 
|\lambda  \sin(2\pi j \beta_l) |$}, so that we obtain  
\begin{equation}
q^{(1)}(\tau) =  \frac{8 \tau}{\pi^2} \cdot  \lim_{n_l \to \infty}
\frac{1}{\ln n_l} \cdot \frac{\displaystyle \sum_{j=1}^{\frac{n_l-1}{2}} \frac{1}{j} 
|\sin(\pi j \beta_l)|^{-\tau} \: |\cos(\pi j \beta_l)|}{\displaystyle 
\frac{2}{n_l} \sum_{j=1}^{\frac{n_l-1}{2}} |\sin(\pi j \beta_l)| ^{-\tau}}. 
\label{equ:q_tau1}
\end{equation} 
To proceed further we have to distinguish between $\tau < 1$ and $\tau >1$.  
For the first case, and provided $\beta=\lim_{l\to+\infty} \beta_l$
is a generic irrational number (not to well approximated by rationals
\cite{Belli1}), we can use the Euler-MacLaurin formula to convert the 
sum in the numerator into  
\begin{equation}
\int_{\beta_{l}}^{\frac{n_l-1}{2} \beta_l} \frac{|\sin(\pi x)|^{-\tau}\: 
|\cos(\pi x)|}{x} \mbox{d}x + C + {\cal O}(\frac{1}{n_l}), 
\end{equation}
where $C$ is a constant. 

Splitting this integral according to the periodicity of its numerator and 
replacing the sum in the denominator of equation (\ref{equ:q_tau1}) by the 
corresponding integral, we finally arrive at the following expression for 
$q^{(1)}(\tau)$:
\begin{equation}
q^{(1)}(\tau) =  \frac{8 \tau}{\pi^2}  \cdot 
\frac{\displaystyle \int_0^1 |\sin (\pi x)|^{-\tau} |\cos (\pi x)| \mbox{d}x}
{\displaystyle \int_0^1 |\sin (\pi x)|^{-\tau} \mbox{d}x}, \quad \tau < 1, 
\end{equation}
which yields 
\begin{equation}
q^{(1)}(\tau) =   \frac{16 }{\pi^{5/2}}  \cdot \frac{\tau \: 
\Gamma(\frac{-\tau+2}{2})}{(-\tau+1)\Gamma(\frac{-\tau+1}{2})}, 
\quad \tau < 1. 
\end{equation}
Therefore we find the following relation for $\tau(q)$ in first order 
in $\lambda$: 
\begin{equation}
\tau + \frac{16}{\pi^{5/2}} |\lambda| 
\frac{\tau \: \Gamma(\frac{-\tau+2}{2})}{(-\tau+1)\Gamma(\frac{-\tau+1}{2})} 
 = q - 1, \qquad \tau < 1.
\label{equ:tau_q}
\end{equation}
Using $\tau(q)=(q-1)D_{q}$ one obtains for the generalized dimensions  
\begin{equation}
D_{q} = 1 - |\lambda| \frac{16}{\pi^{5/2}} 
\frac{\Gamma(\frac{3-q}{2})}{(2-q)\Gamma(\frac{2-q}{2})}, \qquad q < 2, 
\label{equ:D_q}
\end{equation}
which remains valid for $q\to 2$, yielding $D_2 = 1 - \frac{8}{\pi^2}|\lambda|$. 
The limit $\tau \to 1$ corresponds to a limiting point of the graph 
$f(\alpha)$, which is given by:  
\begin{eqnarray}
\alpha_{\rm c}  & = & 1  - \frac{8}{\pi^{2}} |\lambda| \left(1+ \frac{1}{2} 
\left(
\Gamma'(1) - \frac{\Gamma'(\frac{1}{2})}{\sqrt{\pi}} \right) \right),  
\label{equ:a_c}\\
f(\alpha_{\rm c})  & = & 1  - \frac{8}{\pi^{2}} |\lambda| \left( 
1 + \Gamma'(1) - \frac{\Gamma'(\frac{1}{2})}{\sqrt{\pi}} \right). 
\label{equ:f(a_c)}
\end{eqnarray}

The exponent $\gamma$ describing the vanishing of the Lebesgue measure
or total bandwidth ($W\sim n_l^{-\gamma}$)
satisfies $D_{-\gamma} = \frac{1}{1+\gamma}$, and is found to be
$\gamma = 1 - D_0 = \frac{4}{\pi^2}|\lambda|$, in agreement with the result
given in  \cite{Sire}.  As we have made the assumption that $\lambda
q^{(1)}(\tau)$ is small,  the results for  $\tau(q)$ and $D_q$ are not valid
for large negative values of $\tau$ or $q$,  where $q^{(1)}(\tau)$ scales as
$\sqrt{-\tau}$.  Therefore it is not possible to obtain $\alpha_{\rm
max}=D_{-\infty}$  within this approximation.  To check self-consistency of
our results, however, we can set $\alpha_{\rm max}$ equal to the value where
$f(\alpha) = 0$. This yields $\alpha_{\rm max} =  1 - \frac{32}{\pi^{5}}
\lambda^2$, which shows that the decrease of  $\alpha_{\rm max}<1$ is a
second order effect in $\lambda$, in agreement  with the results obtained by the
trace map for  the Fibonacci chain \cite{Ostlund,Kohmoto1,Kohmoto2,Kohmoto3},
and we cannot expect to obtain an exact result for it in this first order
calculation.  We also notice that our result is independent of the irrational
number $\beta$. 
In the case $\tau > 1$, we find that the limit in equation 
(\ref{equ:q_tau1}) vanishes as  $1/\ln n_l$ for $n_l \to \infty$,  
thus yielding $q^{(1)}(\tau) = 0$. 
Therefore the point $(\alpha_{\rm min}=1/2,f(\alpha_{\rm min})=0)$, 
which corresponds to 
the van Hove singularities, does not move in first order of the potential. 
This is also consistent with results based on the trace map for the
Fibonacci chain. 

We have seen that  perturbation theory in first order of the 
quasiperiodic potential leads to multifractal properties for the onsite model 
($\rho = 1,\; \lambda \neq 0$). 
For the off-diagonal model ($\rho \neq1,\; \lambda = 0$), however, 
the half-widths of the gaps $|\Delta E_j| = \frac{2}{\pi j} |1-\rho| 
|\sin(\pi j \beta_l)|$ 
correlates strongly with the corresponding bandwidth of the periodic chain 
$\Delta_{j,0} = \frac{2\pi}{n_l} |\sin(\pi j \beta_l)|$. 
Following the same line of calculations as for the off-diagonal model,
this uniform shrinking of the bandwidths leads to 
\begin{equation}
q^{(1)}(\tau) = \frac{4}{\pi^2}\cdot \tau , \quad \tau <1.
\end{equation}
Therefore the corresponding spectrum is a monofractal with all generalized 
dimensions equal to $D_0 = 1 - \frac{4}{\pi^2}|1-\rho|$. 

We conclude this paragraph with the remark that it is not the feature
diagonal/off-diagonal that leads to the different behavior of the two
models. Indeed,  for the diagonal model $\psi_{i+1} + v_i\psi_i + \psi_{i-1} =
E \psi_i $ with onsite potentials, 
\begin{equation}
    v_{i} = \left\{ \begin{array}{cc}
-\lambda &  \mbox{for}\quad 0<\{\omega i\}<1-\omega, \\
\lambda & \mbox{for}\quad 1-\omega < \{\omega i\} < 1.
                            \end{array}
        \right.
\end{equation}
the band edges are given by \cite{Sire}   
\begin{equation}
E_j \pm \Delta E_j = 2\epsilon \cos(\pi j \beta_l) + \lambda (2\beta_l -1) 
\pm \frac{2\lambda}{\pi j}  \sin(\pi j \beta_l).
\end{equation}
The same argument as above shows that in this case the spectrum is monofractal 
in first order in $\lambda$.

\section{Comparison to numerical data and conclusion}

Numerical calculations have been performed to determine the multifractal 
spectrum. 
In order to speed up the convergence of the numerical data, we calculate 
the multifractal scaling function $\tau(q)$ by using the more
efficient condition,
\begin{equation}
\frac{\Gamma^{(l)}(\tau,q)}{\Gamma^{(l')}(\tau,q)} = 1, 
\label{equ:def_tau_q}
\end{equation}
rather than by using $\Gamma^{(l)}(\tau,q)=1$ \cite{Tang}. 
Furthermore, we correct the systematic error due to the finite size of the 
chains by setting
\begin{equation}
\left. \frac{\Gamma^{(l)}(\tau,q)}{\Gamma^{(l')}(\tau,q)} \right|_{\rm cor}=
\left. \frac{\Gamma^{(l)}(\tau,q)}{\Gamma^{(l')}(\tau,q)} \right|_{\rm num}\cdot 
\left. \frac{\Gamma_0^{(l)}(\tau,q)}{\Gamma_0^{(l')}(\tau,q)} \right|_{\rm ex}\cdot
\left. \frac{\Gamma_0^{(l')}(\tau,q)}{\Gamma_0^{(l)}(\tau,q)} \right|_{\rm num},
\end{equation}
where the subscript ``num'' refers to the data obtained by numerical 
diagonalization and $(\Gamma^{(l)}_0/\Gamma^{(l')}_0)_{\rm ex}$
 is the asymptotically 
exact value for the undisturbed system:
\begin{equation}
\left. \frac{\Gamma_0^{(l)}(\tau,q)}{\Gamma_0^{(l')}(\tau,q)} \right|_{\rm ex}=
\left( \frac{n_l}{n_{l'}} \right)^{q^{(0)}(\tau)-q}. 
\label{powerlaw} 
\end{equation}
This correction procedure is also intended to eliminate the logarithmic 
contribution to the ideal power law (\ref{powerlaw}) expected near 
$(\tau\approx 1,q\approx 2)$, which is already present for the 
periodic chain case
\cite{mosszhong}, and which can strongly affect the numerical determinations
of scaling exponents \cite{mosszhong}.
As we are interested in calculating the first order effects of the quasiperiodic 
potential, we can evaluate $\Gamma^{(l)}(\tau,q)|_{\rm num}$ according 
to equation (\ref{equ:exp_Gamma}), where, of course, $\delta \Delta_j^{(1)}$
has to be replaced by the value $\Delta_{j,0} - \Delta_j |_{\rm num}$. 

Finally the corrected value $(\Gamma^{(l)}/\Gamma^{(l')})|_{\rm cor}$ 
 is inserted into equation 
(\ref{equ:def_tau_q}) for  calculating $\tau(q)$, and, afterwards, $f(\alpha)$
by  Legendre transformation.                    
                                     
\begin{figure}
\vskip0cm
\centerline{\epsfig{figure=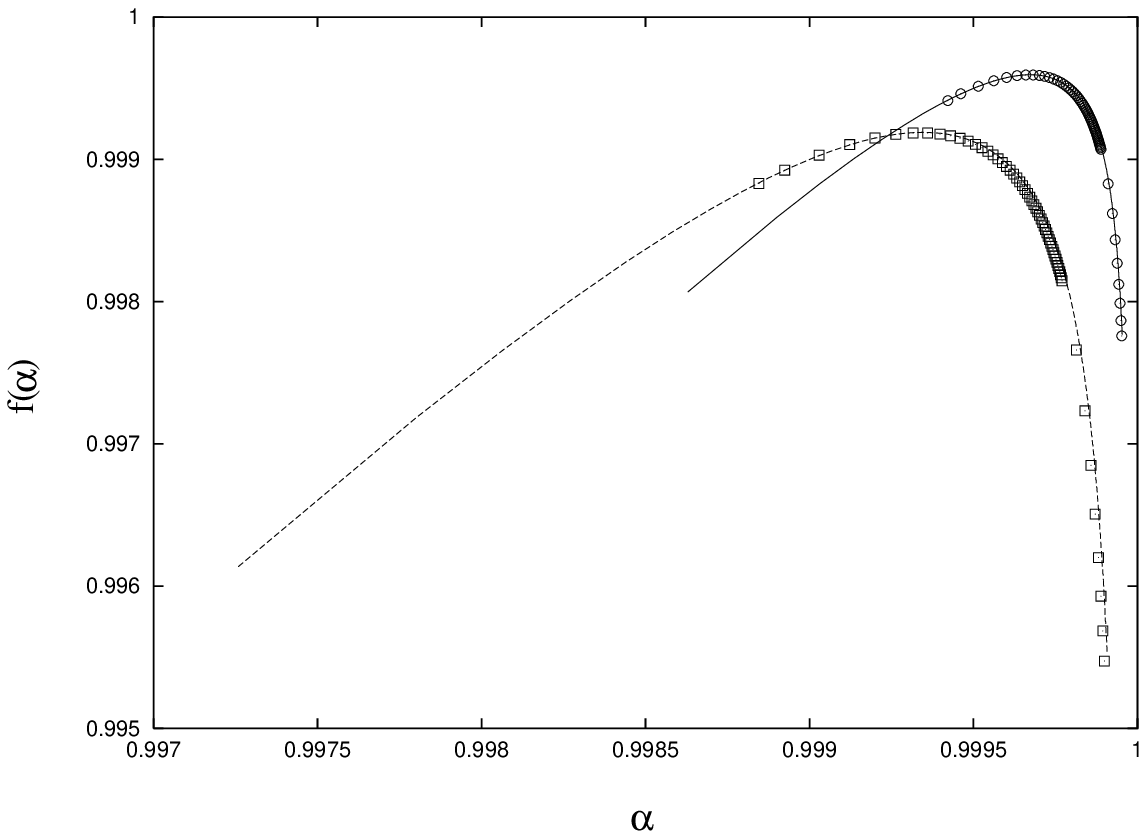,width=12.5cm}}
\caption{The theoretical multifractal spectrum $f(\alpha)$ for the onsite
model ($\rho = 1$) is plotted for  $\lambda=0.001$ (full line) and 
$\lambda=0.002$ (dashed line) and compared to numerical data for the 
Fibonacci chain with $n_l = 987$ and $n_{l'} = 233$
(symbols). The theoretical curves are obtained by Legendre transformation of
equation (21)  in the range $\tau \in (1,50]$. For the numerical data 
we have chosen $\Delta \tau = 0.2$ for $\tau <10$ and  $\Delta \tau =
5$ for $\tau \ge 10$ for sake of clarity.} 
\label{fig:f_alpha}  
\end{figure}

In Fig.\ \ref{fig:f_alpha} the numerical curves $f(\alpha)$ are compared with 
the theoretical expression for the onsite model and two different values of 
$\lambda$ ($\lambda=0.001,\, \lambda=0.002$).                                          
The limiting points at the left side correspond to $\tau \to 1$ and are given 
by the expressions (\ref{equ:a_c}) and (\ref{equ:f(a_c)}). 
The points calculated from the numerical data are in good agreement
with the theoretical result. 
The difficulty of approaching the limiting point $(\alpha_{\rm c},f(\alpha_{\rm
c}))$ is due to the very slow convergence of the limit of equation
(\ref{equ:q_tau1}), when $\tau$ approaches 1. 
As we have seen, $q^{(1)}(\tau)$ shows a discontinuity for $\tau = 1$,  thus
preventing good convergence for a finite system even quite far from the critical
value.  This explains why the numerical points end before reaching $\alpha =
\alpha_{\rm c}$. However, the numerically accessible $\alpha_c$ 
systematically approaches the theoretical value as the
system size increases. 
 
\begin{figure}
\centerline{\epsfig{figure=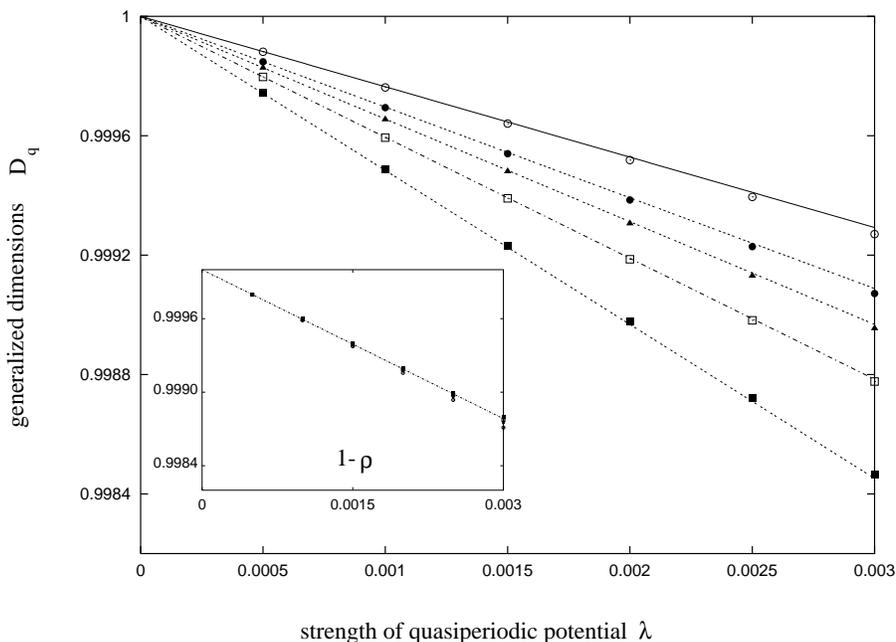,width=12.5cm}}
\caption{The theoretical dependence of the generalized dimensions $D_q$ on the
quasiperiodic  potential $\lambda$ for the onsite model, for $q=-5,-2,-1,0,1$
(dashed lines from top to bottom) is compared to 
numerical data (symbols). Insert:
same as before for the off-diagonal model ($\lambda = 0$).} 
\label{fig:D_q}
\end{figure}

Fig.\ \ref{fig:D_q} shows the dependence of the generalized 
dimensions $D_{q}$ ($q=-5,-2,-1,0,1$) on the quasiperiodic potential 
$\lambda$ for the onsite model.   
The lines are the theoretical expressions (equation (\ref{equ:D_q})). 
Comparison with the insert, where the corresponding data 
are plotted for the off-diagonal model, shows that the  different behavior
of the two models that has been found theoretically  is confirmed by the 
numerical calculations. 

In conclusion, we have calculated the multifractal spectrum for a general 
quasiperiodic chain in first order in the quasiperiodic potential. 
We have found analytical expressions for the generalized dimensions which are 
in good agreement with numerical calculations. 
The two models under investigation display qualitatively different behavior, 
the onsite model showing multifractal behavior, the off-diagonal being 
monofractal. 
For both models the multifractal properties do not depend on the irrational 
slope in first order in the potential. 

\vskip 0.4cm
{\it Acknowledgements:} This work was supported by the EC under the 
network reference ERBCHRXCT940528; one of us (A.R.) would like to thank the 
Laboratoire de Physique Quantique (Toulouse) for its kind hospitality.

\begin{references}
\bibitem{remark} Strictly speaking perturbation theory yields $n_{l}\sin(\pi
  j/n_{l})$ instead of $\pi j$ in the denominator of equation
  (\ref{equ:start}) (cf. equation (10) in \cite{Sire}). Our equation
  (\ref{equ:start}) corresponds to equation (17) in \cite{Sire}; it can be
  shown that the scaling behavior, in which we are interested, is not affected
  by the approximation employed.
\end{references}
\end{document}